\documentclass{PoS}
\usepackage{subfig}
\newcommand{\rmd}{\textrm{d}}
\newcommand{\Fermi}{\emph{Fermi-LAT}}
\title{Dark Matter searches with H.E.S.S. towards dwarf spheroidal galaxies}

\ShortTitle{Dark Matter searches with H.E.S.S. towards dwarf spheroidal galaxies}

\author{\speaker{A. Viana} on behalf of the H.E.S.S. collaboration\\
        CEA - Saclay, DSM/Irfu/SPP,
Gif-sur-Yvette, 91191, France \\
        E-mail: \email{aion.viana@cea.fr}}

\abstract{The H.E.S.S. experiment is an array of four identical imaging atmospheric Cherenkov telescopes in the Southern hemisphere, designed to observe very high energy gamma-rays (E > 100 GeV). These high energy gamma-rays can be used to search for annihilations of Dark Matter particles in dense environments. Dwarf galaxy dynamics shows that they are Dark Matter-dominated environments. Several observation campaigns on dwarf satellite galaxies of the Milky Way were launched by H.E.S.S.. The observations are reviewed. In the absence of clear signals, constraints on the Dark Matter particle annihilation cross-section have been derived in different particle physics scenarios. Some possible enhancements of the gamma-ray flux are studied, \textit{i.e.}, the Sommerfeld effect, the internal bremsstrahlung and the substructures in the Dark Matter halo.}

\FullConference{Identification of Dark Matter 2010-IDM2010\\
		July 26-30, 2010\\
		Montpellier France}

\begin{document}

\section{Introduction}
Over the past decades, compelling evidences have been accumulated suggesting
a sizeable non-baryonic dark matter (DM) component in the total cosmological
energy density of the Universe. The present estimate of the cold DM density is
$\rm \Omega_{CDM}h^2 \simeq $ 0.111 $\pm$ 0.006 where the scaled Hubble parameter is h $\simeq$ 0.70.
At the galactic scales, evidences come from the measurements of the rotation curves for
spiral galaxies as well as the gravitational lensing, and agree well with the predictions
of N-body simulations of gravitational clustering in the CDM cosmology. At higher
scales, the velocity dispersion of galaxies in galaxy cluster suggests high mass-to-light
ratio, exceeding by at least one order of magnitude the ratio in the solar
neighborhood. However, the intrinsic nature of the DM particle remains unknown.
Scenarios beyond the Standard Model of particle physics predict plausible WIMP (weakly interacting
massive particle) candidates to account for the Cold Dark Matter. Among these are the minimal supersymmetric
extension of the Standard Model (MSSM) or universal extra dimension (UED) theories. With R-parity
conservation, SUSY models predict the lightest SUSY particle (LSP) to be stable. In various SUSY
breaking scenarios, the LSP is the lightest neutralino $\widetilde{\chi}$. In Kaluza-Klein (KK) models with KK-parity conservation, the lightest KK particle (LKP) is stable~\cite{Servant},
the most promising being the first KK mode of the hypercharge gauge boson, $\widetilde{B}^{(1)}$.
The annihilation of WIMPs in galactic halos may produce gamma-ray signals
detectable with Cherenkov telescopes (for a review, see~\cite{Bertone}). Generally, their annihilations will lead
to a continuum of gamma-rays with a energy cut-off at the WIMP mass,
from the hadronization and decay of the cascading annihilation products.

The dwarf spheroidal galaxies (dSphs) of the Local Group are the most common
satellites of the Milky Way and assumed to be gravitationally bound
dominantly by Dark Matter (DM). Although the predicted very high energy
(VHE, $E \gtrsim 100$~GeV) gamma ray flux from DM annihilation from dwarf
galaxies is smaller compared to the expected DM annihilation gamma ray
flux from denser regions of DM such as the Galactic Center, these
galaxies are promising targets for searches for gamma rays from DM
annihilation since they are environments with a favorably
low astrophysical gamma ray background. The galaxies themselves are
expected to contain little or no astrophysical gamma ray sources since no recent
star formation activity gives rise to VHE gamma rays (supernova
remnants, pulsar wind nebula, etc.) and little gas acting as
target material for cosmic rays has been measured. 

\section{Search for VHE gamma rays from dwarf spheroidal galaxies by H.E.S.S.}

\subsection{The H.E.S.S. instrument}
The H.E.S.S. (High Energy Stereoscopic System) experiment is an array of four identical imaging atmospheric Cherenkov telescopes, observing VHE gamma rays. H.E.S.S. is located in the Khomas Highland of Namibia ($23^{\circ} 16^{\prime} 18^{\prime \prime}$ South, $16^{\circ}30^{\prime}00^{\prime \prime}$ East) at an altitude of 1800~m above sea level. Each telescope has an optical reflector consisting of 382 round facets of 60~cm diameter each, yielding a total mirror area of 107~m$^2$.
 The Cherenkov light, emitted by charged particles in the electromagnetic showers initiated by primary gamma rays, is focused on cameras equipped with 960 photomultiplier tubes, each one subtending a field-of-view of $0.16^\circ$.
 The large field-of-view ($\sim$$5^\circ$) permits survey coverage in a single pointing. The direction and the energy of the primary gamma rays are reconstructed by the stereoscopic technique~\cite{2004APh....22..285F}.

\subsection{Nearby dwarf galaxies: Sagittarius and Canis Major}

In June and November 2006 H.E.S.S. has observed the Sagittarius (Sgr) dSph (distance $\sim 24$ kpc~\cite{1998ARA&A..36..435M}) and the overdensity Canis Major (CMa) (distance $\sim 8$ kpc~\cite{1998ARA&A..36..435M}), respectively.
The observations of Sgr dSph was taken with an averaged zenith angle of observation of 19$^{\circ}$ and a total of 11h of high quality data are available after data selection. There is no evidence for a VHE gamma ray signal above the energy threshold at the target position~\cite{Sag}. The nature of CMa is still controversial and one scenario represents it as a dwarf galaxy. The observations of CMa were carried out with a zenith angle of observation close to the zenith and extending up to 20$^{\circ}$. A total of 9.6 hours of high quality data were collected and no evidence for a VHE gamma ray signal is found~\cite{Canis}.
Since no significant gamma ray excess was found above the estimated backgrounds
at the nominal positions of Sgr and CMa dSphs, 95\% confidence level upper limits on the total observed numbers of gamma-rays were derived. All the limits have been computed using the method described in Feldman and Cousins~\cite{1998PhRvD..57.3873F}.

\subsection{Distant dwarf galaxies: Sculptor and Carina}

Because they are located at large distances from the sun, the flux attenuation due to the distance is much important in distant dwarf galaxies such as Sculptor (distance $\sim 79$ kpc~\cite{1998ARA&A..36..435M}) and Carina (distance $\sim 101$ kpc~\cite{1998ARA&A..36..435M}) dSphs, which are among the most luminous dSphs satellites of the Milky Way. Nevertheless, this same large distance guarantees less tidal disruption by the the Milky Way, leading to less uncertainties for the DM content of the distant galaxies than the one of the nearby galaxies.
Observation campaigns were launched by H.E.S.S. between January 2008 and December 2009 for the Sculptor and Carina dSphs. The data used for the analysis were taken at average zenith angles of $\sim$$14^\circ$ and $\sim$$34^\circ$ for the Sculptor and Carina dSphs, respectively~\cite{ScuCar}.
The total data set amounts to 11.8 h for Sculptor and 14.8 h for Carina of live time after the quality selection.
Since no significant gamma ray excess was found above the estimated backgrounds at the nominal positions of Sculptor and Carina dSphs, 95\% confidence level upper limits on the total observed numbers of gamma-rays were derived.

\section{Exclusion limits for Dark Matter annihilations}
The gamma ray flux from the annihilations of DM particles of mass $m_{\rm DM}$ in a DM halo is given by a particle physics term times an astrophysics term:
\begin{equation}
\label{eqnp}
\frac{\rmd\Phi_{\gamma}(\Delta\Omega,E_{\gamma})}{\rmd E_{\gamma}}\,=\frac{1}{8\pi}\,\underbrace{\frac{\langle
\sigma v\rangle}{m^2_{\rm DM}}\,\frac{\rmd N_{\gamma}}{\rmd E_{\gamma}}}_{\rm Particle\,
Physics}\,\times\,\underbrace{\bar{J}(\Delta\Omega)\Delta\Omega}_{\rm Astrophysics} \, ,
\end{equation}
where the astrophysical factor $\overline{J}$ is defined as
\begin{equation}
\overline{J}(\Delta\Omega) = \frac{1}{\Delta \Omega} \int_{\Delta \Omega} \rmd\Omega \int_{\rm los} \rho^2[r(s)] \rmd s \, .
\label{jbar}
\end{equation}
In Eq. (~\ref{jbar}) the squared density of DM ($\rho^2$) is integrated along the line of sight (los) and over the solid angle $\Delta\Omega $. The solid angle is fixed to match the angular resolution of the telescope for a point-like source search. For the H.E.S.S. experiment $\Delta \Omega = 10^{-5}$~sr. The particle physics term contains the DM particle mass, $m_{\rm DM}$, the velocity-weighted annihilation cross section, $\langle \sigma v\rangle$, and the differential gamma ray spectrum from all final states weighted by their corresponding branching ratios, $\rmd N_{\gamma}/\rmd E_{\gamma}$.

In order to calculate the exclusion limits on the DM annihilation cross section, one needs to model the density distribution of DM in the observed target that will be used in the astrophysical factor $\overline{J}$ calculation. Two hypotheses for spherical DM halo profiles are used for Sagittarius, Canis Major, Sculptor and Carina: a cored profile~\cite{Swaters:2000nt,Evans2004}, and the \emph{Navarro, Frenk, and White} (NFW) profile~\cite{1996ApJ...462..563N}. The $95\%$ C.L. upper limit on the velocity-weighted annihilation cross section as function of the DM particle mass for a given halo profile is given by
\begin{equation}
\left\langle \sigma v \right\rangle_{\rm min}^{95\%\,{\rm C.L.}} = \frac{8\pi}{\overline{J}(\Delta \Omega)\Delta \Omega} \times \frac{m_{\rm DM}^{2}\, N_{\gamma,\, {\rm tot}}^{95\%\,{\rm C.L.}}}{T_{\rm obs}\, \int_{0}^{m_{\rm DM}} A_{\rm eff}(E_{\gamma}) \, \frac{\rmd N_{\gamma}}{\rmd E_{\gamma}}(E_\gamma) \, \rmd E_{\gamma}} \, ,
\end{equation}
where the parametrization of $\rmd N_{\gamma}$/$\rmd E_{\gamma}$ is taken from~\cite{1998APh.....9..137B} for neutralino self-annihilation, and calculated from~\cite{Servant} for Kaluza-Klein $\widetilde{B}^{(1)}$ self-annihilation.

\begin{figure}
  \centering
  \label{fig:exclusionSc}\includegraphics[scale=0.25]{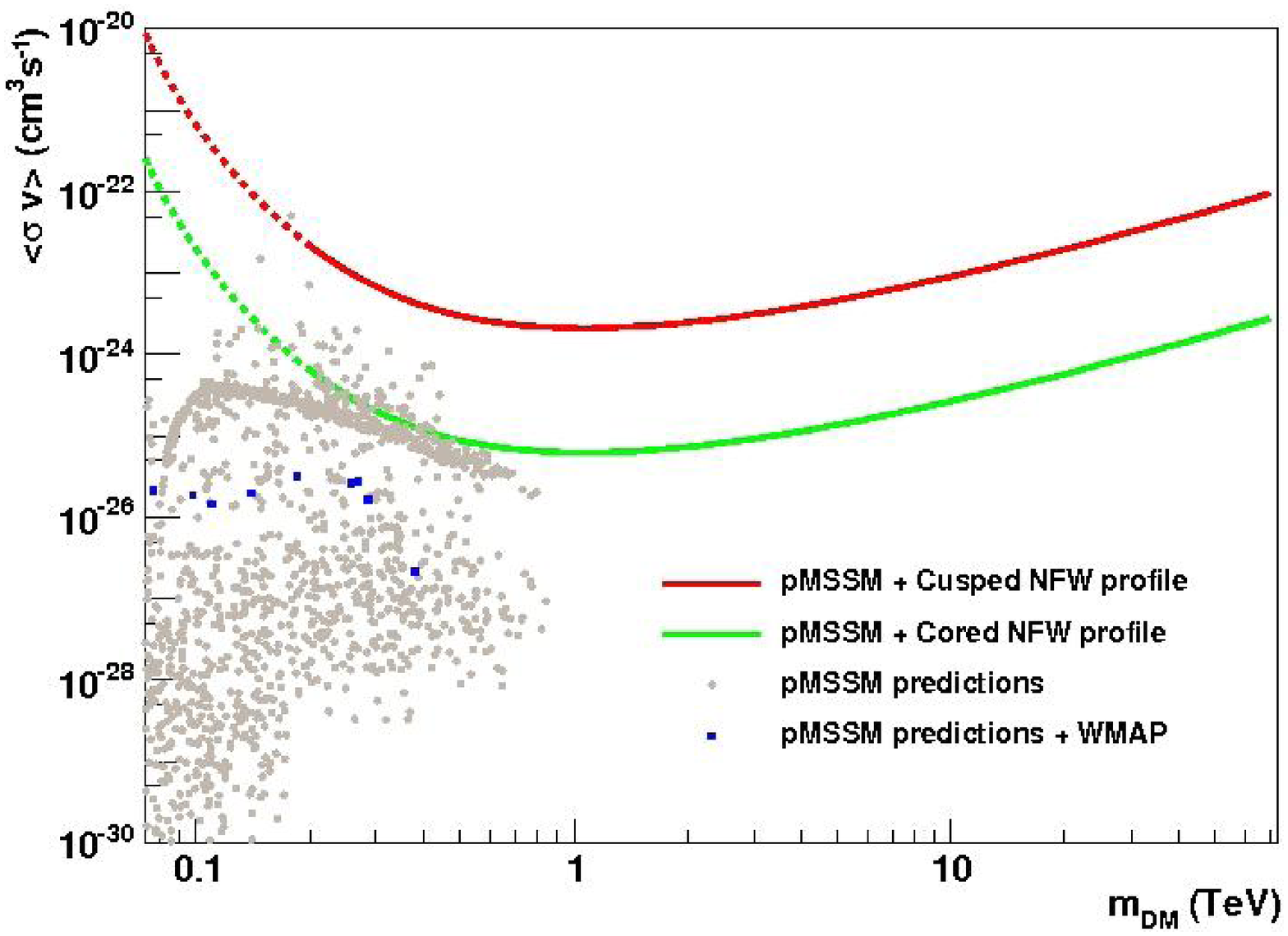}
  \label{fig:exclusionCa}\includegraphics[scale=0.25]{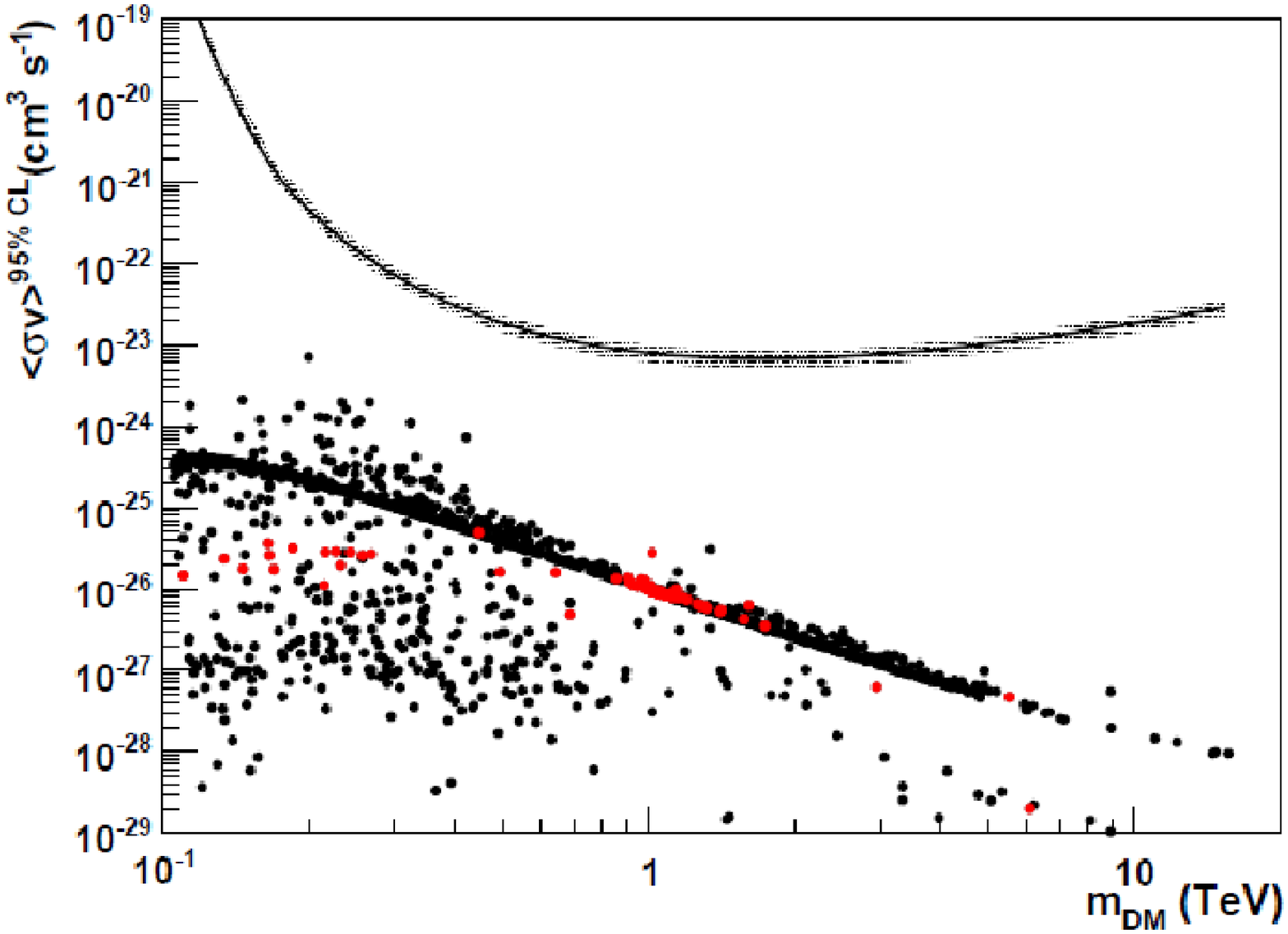}
  \caption{Upper limit at $95\%$ C.L. of $\left\langle \sigma v \right\rangle$ as function of the DM particle mass for a NFW (red line) and a cored (green line) DM halo profiles, respectively, for Sagittarius (\emph{left}) and for a NFW DM halo profiles in the case of Canis Major (\emph{right}) (see~\cite{Sag} and~\cite{Canis} for more details).
  \label{fig:exclusionSC}}
\end{figure}

\begin{figure}
  \centering
  \label{fig:exclusionSc}\includegraphics[scale=0.3]{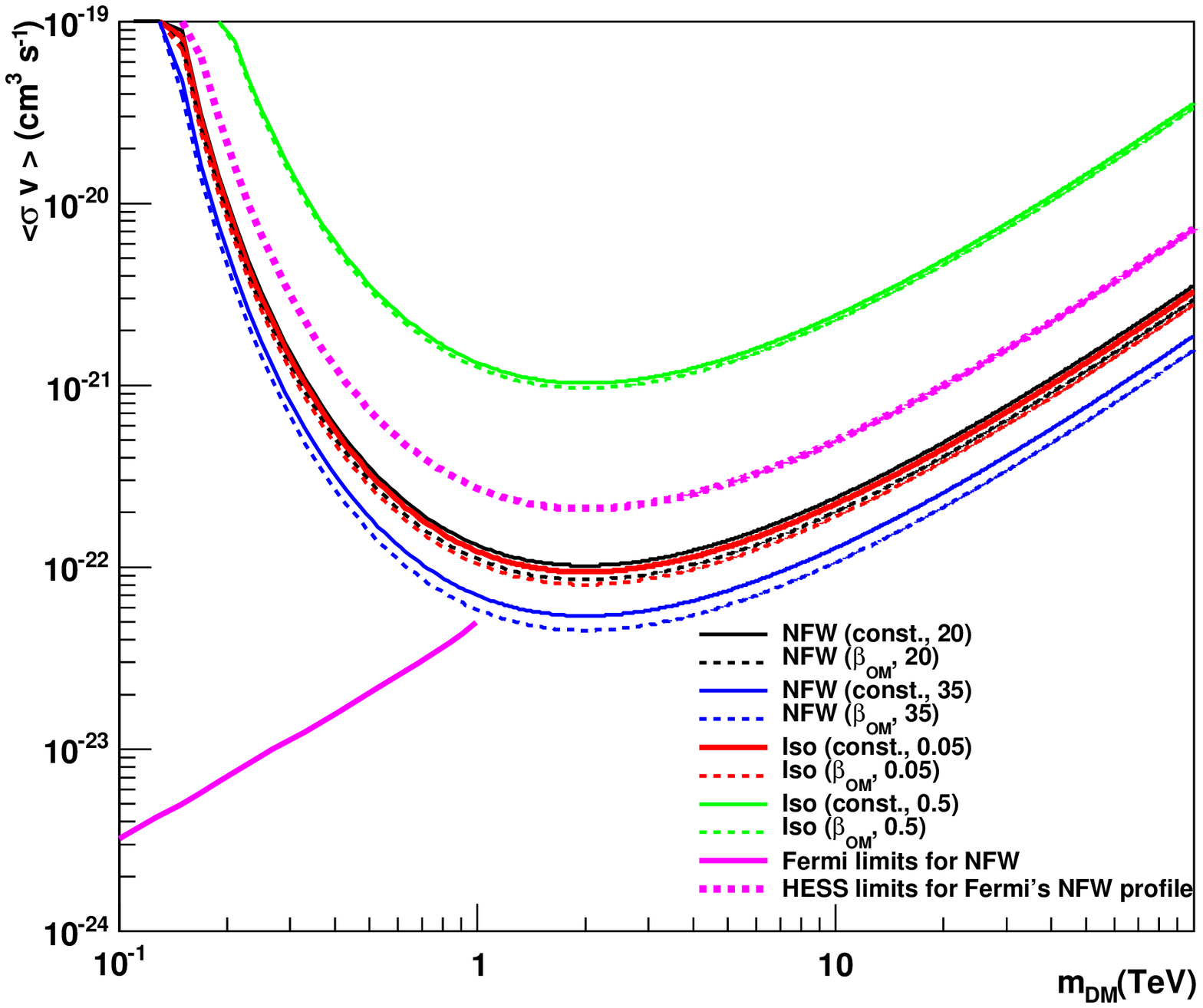}
  \label{fig:exclusionCa}\includegraphics[scale=0.3]{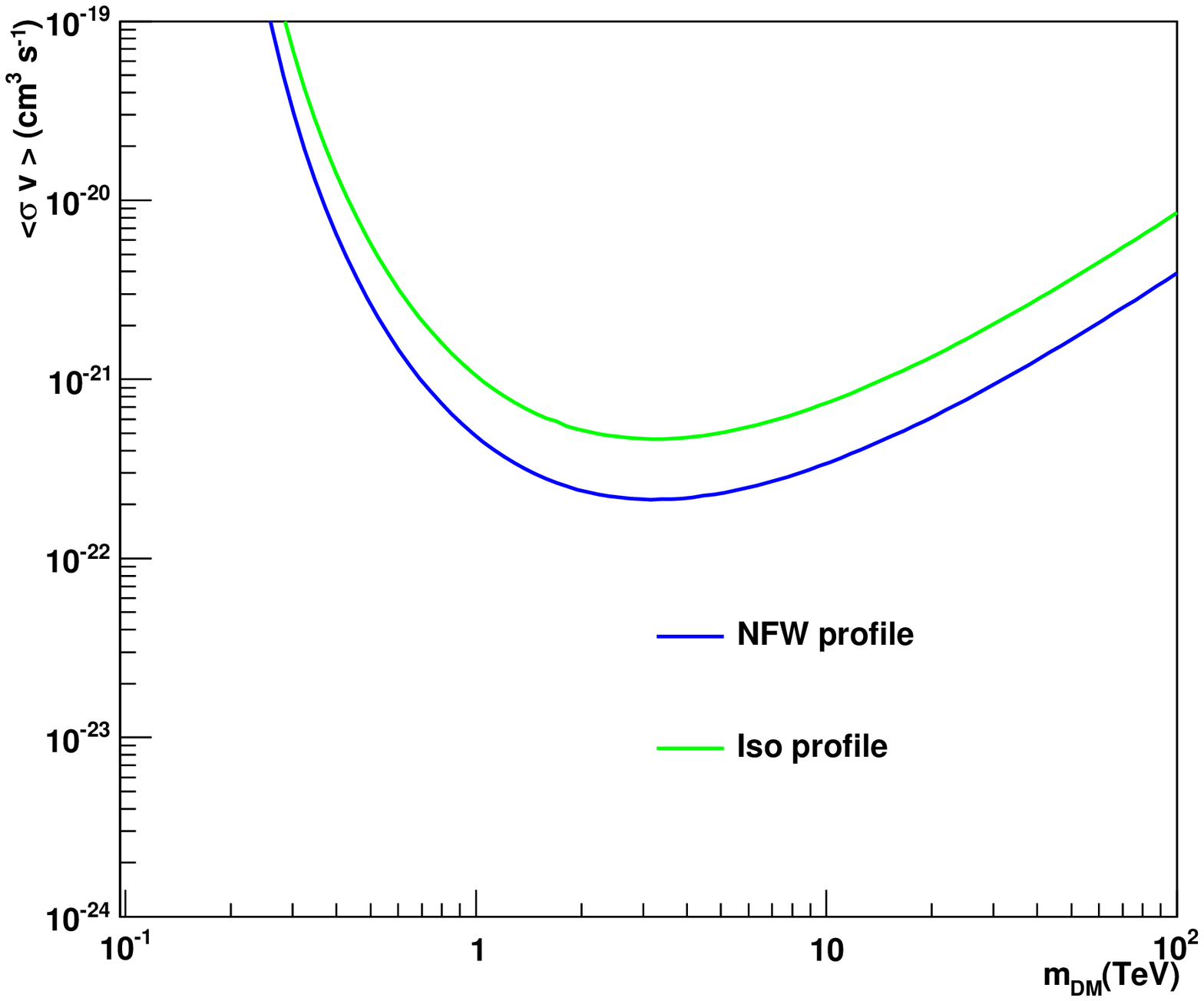}
  \caption{Upper limit at $95\%$ C.L. of $\left\langle \sigma v \right\rangle$ as function of the DM particle mass for several  NFW halo profiles and the pseudo-isothermal (cored) DM halo profiles for Sculptor (\emph{left}) and Carina (\emph{right}), see~\cite{ScuCar} for more details. The \Fermi\ limits~\cite{2010ApJ...712..147A} for a NFW profile are also plotted as well as the H.E.S.S. limits for this NFW profile (pink dashed line).
   \label{fig:exclusion}}
\end{figure}

The exclusion curves for the neutralino case are plotted for the Sagittarius and Canis Major dSphs in Figure~\ref{fig:exclusionSC} and the ones for the Sculptor and Carina dSphs are plotted in Figure~\ref{fig:exclusion} referring to several DM halo profiles parametrizations. The \Fermi\ exclusion limit for Sculptor is also plotted~\cite{2010ApJ...712..147A} for a NFW profile and a neutralino self-annihilation spectrum in $b\overline{b}$ final state. The H.E.S.S. exclusion limits are also calculated using the NFW profile from \Fermi\ \cite{2010ApJ...712..147A}.
The exclusion curves for the Kaluza-Klein DM particle $\widetilde{B}^{(1)}$ case can be found in~\cite{Sag} and~\cite{Canis} for Sagittarius and Canis Major, respectively, and in~\cite{ScuCar} for Sculptor and Carina.

\section{Enhancement effects of the gamma ray flux}

 For Sculptor and Carina dSphs three cases that can modify their exclusion limits are considered: two particle physics effects, namely the \emph{Sommerfeld enhancement} and the \emph{Internal Brems\-strahlung} (IB) from the DM annihilation, and an astrophysical effect due to the mass distribution of \emph{dark-matter substructures}.

\subsection{The Sommerfeld enhancement}

Here two new assumptions are made for the Sculptor's and Carina's DM halo composition. The first assumption is to assume the DM particle annihilates into a pair of W boson, which is the case when the neutralino is a pure wino. The second assumption is to assume that the DM mean velocity dispersion inside the halo is the same as for the stars (the mean velocity dispersion of the stars is $\sigma_{\rm V}\sim 10.0$~km/s for Sculptor and $\sigma_{\rm V}\sim 7.5$~km/s for Carina).

In this class of objects, the relative velocity between the DM particles may be sufficiently low so that the Sommerfeld effect can substantially boost the annihilation cross section~\cite{Lattanzi:2008qa}, since it is particulary effective in the very low-velocity regime. This non-relativistic effect arises when two DM particles interact in an attractive potential. The actual velocity-weighted annihilation cross section of the neutralino can then be enhanced by a factor S defined as
\begin{equation}
\left\langle \sigma v \right\rangle = S\left\langle \sigma v \right\rangle_0 \, ,
\end{equation}
where the value of S depends on the DM particles mass and mean velocity dispersion, the exchange boson mass and coupling.
A wino would interact with the attractive potential created by the Z gauge boson through the
weak force before annihilation occurs, which would give rise to an enhancement. The value of this enhancement was numerically
calculated as done in~\cite{Lattanzi:2008qa} and then used to improve the $95\%$~C.L. upper limit on the velocity-weighted annihilation cross section, $\left\langle \sigma v \right\rangle / S$ as a function of the DM particle mass.

\subsection{Internal Bremsstrahlung}

In some specific regions of the MSSM parameter space, the electromagnetic radiative correction to the main annihilation channels into charged particles can give a significant enhancement to the expected gamma ray flux in the observed environment due to internal Bremsstrahlung (IB)~\cite{Bergstrom:1989jr,Bringmann:2007nk}. In the \emph{stau co-annihilation region} of the minimal supergravity (mSUGRA) models, for instance, the wino annihilation spectrum would receive a considerable contribution from the internal Bremsstrahlung.

\begin{figure}[h!]
  \begin{center}
    \mbox{\hspace{0cm}\includegraphics[scale=0.43]{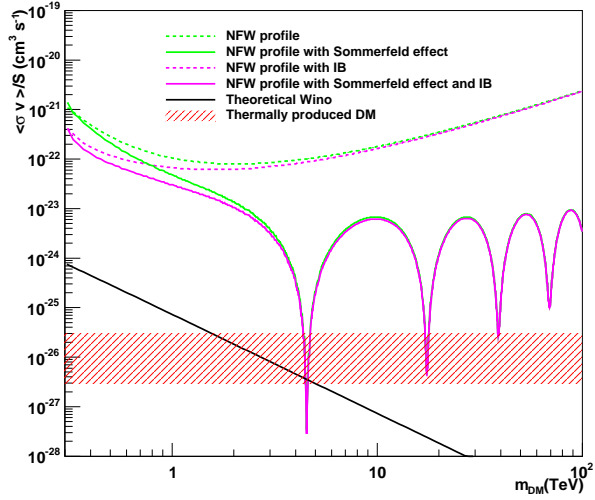}}
    \caption{Upper limit at $95\%$ C.L. on $\left\langle \sigma v \right\rangle /S $ as function of the DM particle mass enhanced by the Sommerfeld effect and the internal Bremsstrahlung  (see text for more details)  for a NFW profile of Sculptor. The predicted $\left\langle \sigma v \right\rangle_0$ for a pure wino (solid black line) as well as the typical cross section for a thermally produced DM (dashed red area) are also plotted.\label{fig:exclusionSom_IB}}
  \end{center}

\end{figure}

 This contribution to the annihilation spectrum was computed using the parametrization of~\cite{Bringmann:2007nk} for all the wino masses in the H.E.S.S. energy range. The enhancement effect on the $95\%$~C.L. upper limit on the velocity-weighted annihilation cross section is shown in Figure~\ref{fig:exclusionSom_IB}. The joint enhancement due to the Sommerfeld effect and IB is also plotted. The predicted $\left\langle \sigma v \right\rangle_0$ for a pure wino as well as the typical annihilation cross section for a thermally produced DM ($\left\langle \sigma v \right\rangle_0$ $\sim 10^{-26}$~cm$^{3}$s$^{-1}$~\cite{1996PhR...267..195J}) are also plotted. Some specific wino masses can be excluded at the level of $\left\langle \sigma v \right\rangle_0$ $\sim 10^{-26}$~cm$^{3}$s$^{-1}$. The effect of the IB is only significant in the exclusion limits for the low mass DM particle regime.

\subsection{Enhancement from dark-matter substructures}

Astrophysical effects may also modify the exclusion limits. Numerical simulations of galactic halos predict a population of substructures that could contribute to the overall astrophysical factor in Eq. (~\ref{jbar}). Using the procedures given in~\cite{2009MNRAS.399.2033P}, the contribution to the astrophysical factor by the DM substructures population was estimated for Sculptor and Carina. An enhancement of the astrophysical factor is found to be of a few percent, which is too small to significantly affect the exclusion limits presented.

\section{Conclusions}

The H.E.S.S. collaboration has observed several dwarf spheroidal galaxies and searched for DM annihilations. The first round of observation in 2006 encompassed the nearby dwarf galaxies Sagittarius and Canis Major. Subsequently in 2008 a second round of observations was made on Sculptor and Carina dSph. No significant gamma ray excess was found at the nominal target position of these galaxies.

Constraints have been obtained for the velocity weighted annihilation cross section $\left\langle \sigma v \right\rangle$ as a function of the mass for neutralino and KK DM particles. H.E.S.S. limits are comparable to the limits reported by MAGIC~\cite{Albert:2007xg} and VERITAS~\cite{Acciari:2010pja}, but weaker than those obtained by \Fermi~\cite{2010ApJ...712..147A} in the GeV mass range. Nevertheless, they are complementary to the \Fermi\, limits in the TeV range. Finally assuming a resonance effect in the Sommerfeld enhancement, some specific wino masses can be excluded, and the first experimental constraints have been obtained on the Sommerfeld effect using H.E.S.S. data and DM annihilation spectra.
Dark matter searches will continue and searches with H.E.S.S. 2 will start in 2012. The phase 2 will consist
of a new large 28 m diameter telescope located at the center of the existing array. With the availability
of the large central telescope H.E.S.S. 2, the analysis energy threshold will be lowered down to less than 80
GeV and will allow to explore a lower DM masse range.

\end{document}